# The discrete dipole approximation for simulation of light scattering by particles much larger than the wavelength

M.A. Yurkin[a,b*], V.P. Maltsev[b,c], and A.G. Hoekstra[a*]

[a] *Section Computational Science, Faculty of Science, University of Amsterdam, Kruislaan 403, 1098 SJ, Amsterdam, The Netherlands*
[b] *Institute of Chemical Kinetics and Combustion, Siberian Branch of the Russian Academy of Sciences, Institutskaya Str. 3, 630090, Novosibirsk, Russia*
[c] *Novosibirsk State University, Pirogova Str. 2, 630090, Novosibirsk, Russia*

## Abstract

In this manuscript we investigate the capabilities of the Discrete Dipole Approximation (DDA) to simulate scattering from particles that are much larger than the wavelength of the incident light, and describe an optimized publicly available DDA computer program that processes the large number of dipoles required for such simulations. Numerical simulations of light scattering by spheres with size parameters $x$ up to 160 and 40 for refractive index $m = 1.05$ and 2 respectively are presented and compared with exact results of the Mie theory. Errors of both integral and angle-resolved scattering quantities generally increase with $m$ and show no systematic dependence on $x$. Computational times increase steeply with both $x$ and $m$, reaching values of more than 2 weeks on a cluster of 64 processors. The main distinctive feature of the computer program is the ability to parallelize a single DDA simulation over a cluster of computers, which allows it to simulate light scattering by very large particles, like the ones that are considered in this manuscript. Current limitations and possible ways for improvement are discussed.

---

[*] Corresponding authors. Tel.: +31-20-525-7462; fax: +31-20-525-7490.
*e-mail addresses*: myurkin@science.uva.nl, alfons@science.uva.nl

# 1 Introduction

The discrete dipole approximation (DDA) is a general method to calculate scattering and absorption of electromagnetic waves by particles of arbitrary geometry and composition. The DDA was first proposed by Purcell and Pennypacker [1] and was reviewed by Draine and Flatau in 1994 [2]. A recent review [3] describes the current state of the DDA and its historical development. It also explains the equivalence of the DDA and methods based on the volume integral equation formulation. The reader is referred to this review for an in-depth discussion of the DDA.

There are a number of computer programs based on the DDA, some of which were recently compared by Penttila *et al.* [4]. The most popular among them is DDSCAT [5], which has been widely used by many researchers for more than 10 years. In this paper we present a new program, Amsterdam DDA (ADDA), which recently has been put in the public domain.[1] Its main distinctive feature is the ability to parallelize a single DDA simulation over a cluster of computers, which allows simulation of light scattering by very large particles. This is demonstrated for a number of test cases in this manuscript. Validation of ADDA by simulating light scattering by wavelength-sized particles and comparing it with other DDA programs was reported elsewhere [4].

Section 2 describes in detail the ADDA computer code, showing its advantages compared to other codes. A number of numerical tests are shown in Section 3, demonstrating that DDA is actually capable processing large particles, and showing the current capabilities of ADDA. Results of these simulations are discussed in Section 4; the errors are compared with previous results for much smaller particles. Section 5 concludes the manuscript and discusses possible future work.

# 2 ADDA computer code

ADDA has been developed over a period of more than 10 years at the University of Amsterdam [6-8]. Its main feature (distinctive from other DDA codes) has always been the capability of running on a cluster of computers, parallelizing a *single* DDA computation, in contrast with e.g. DDSCAT [5] that allows farming several instantiations of a DDA simulation to different processors. This allows using a practically unlimited number of dipoles, since ADDA is not limited by the memory of a single computer [8,9]. Recently the overall performance of the code has been improved significantly, together with some optimizations specifically for single-processor mode. ADDA's source code and documentation is freely available.

Most of ADDA is written in ANSI C, which ensures wide portability on the source-code level. The code is fully operational under Linux and, in sequential mode, on Windows based systems. The parallelization over multiple processors is based on a geometric decomposition of the particle and the single-program-multiple-data paradigm of parallel computing. The code is written for distributed memory systems using the message passing interface (MPI).[2] Note that ADDA should in principle also run on shared memory computers, but so far this was not explicitly tested. The fast Fourier transform (FFT) used for the matrix-vector products in the iterative solver is performed either using routines by Temperton [10] or the more advanced package "Fastest Fourier transform in the West" (FFTW) [11]. The latter is generally considerably faster but requires a separate package installation.

ADDA has four options implemented for dipole polarizabilities: Clausius-Mossotti [1], radiative reaction correction [12], lattice dispersion relation (LDR) [13], and corrected LDR [14]. It includes four iterative methods: conjugate gradient applied to normalized equation with minimization of residual norm (CGNR) [15], Bi-CG stabilized (Bi-CGSTAB) [15], Bi-

---
[1] http://www.science.uva.nl/research/scs/Software/adda/
[2] http://www.mpi-forum.org

CG [16], and quasi minimal residual (QMR) [16]. The last two iterative methods employ the complex-symmetric property of the DDA interaction matrix to halve the calculation time [16]. The default stopping criterion of the iterative method in ADDA is the relative norm of the residual $\varepsilon$, which must be $<10^{-5}$.

The usual formulation of DDA can be written as [2,3]:

$$\bar{\alpha}_i^{-1} \mathbf{P}_i - \sum_{j \neq i} \bar{\mathbf{G}}_{ij} \mathbf{P}_j = \mathbf{E}_i^{inc}, \quad (1)$$

where $\bar{\alpha}_i$ is the tensor of dipole polarizability, $\mathbf{E}_i^{inc}$ is incident electric field, $\bar{\mathbf{G}}_{ij}$ is the free-space Green's tensor (complex symmetric), and $\mathbf{P}_i$ is the unknown dipole polarization. If the polarizability tensor is diagonal for all dipoles then there always exists a $\bar{\beta}_i$ such that $\bar{\beta}_i \bar{\beta}_i = \bar{\alpha}_i$, i.e. $\bar{\beta}_i = \sqrt{\bar{\alpha}_i}$. Moreover, $\bar{\beta}_i$ is then complex symmetric, and so is the matrix with elements

$$\bar{\mathbf{A}}_{ij} = \begin{cases} \bar{\mathbf{I}}, i = j \\ -\bar{\beta}_i \bar{\mathbf{G}}_{ij} \bar{\beta}_j, i \neq j \end{cases}, \quad (2)$$

where $\bar{\mathbf{I}}$ is an identity tensor. $\bar{\mathbf{A}}$ is the interaction matrix that is used in ADDA, i.e. the following system of linear equations is solved:

$$\sum_j \bar{\mathbf{A}}_{ij} \mathbf{x}_j = \mathbf{x}_i - \sum_{j \neq i} \bar{\beta}_i \bar{\mathbf{G}}_{ij} \bar{\beta}_j \mathbf{x}_j = \bar{\beta}_i \mathbf{E}_i^{inc}, \quad (3)$$

where $\mathbf{x}_i = \bar{\beta}_i^{-1} \mathbf{P}_i$ is a new unknown vector. Eq. (3) is equivalent to the use of Jacobi-preconditioning [15] together with keeping the interaction matrix complex-symmetric (for any distribution of refractive index inside the scatterer and for any of the supported polarization prescriptions). We have not studied, however, whether this Jacobi-preconditioning improves the convergence of the iterative solver. Flatau showed [17] that in some test cases it helps, while in others there is no improvement. It is important to note also that DDA is not limited to diagonal or symmetric polarizabilities. Any other tensor may be used, but then the interaction matrix is not complex-symmetric; hence, QMR and Bi-CG are less efficient.

ADDA can perform orientation averaging of the scattering quantities over three Euler angles ($\alpha$, $\beta$, $\gamma$) of the particle orientation. Averaging over the angle $\alpha$ is done with a single computation of internal fields by computing scattering in different scattering planes, which is comparably fast. Averaging over the other two Euler angles is done by independent DDA simulations. The averaging itself is performed using a Romberg integration [18], which may be used adaptively (i.e. automatically simulating the required number of different orientations to reach a prescribed accuracy) but limits the possible number of values for each orientation angle to be $2^n + 1$, where $n$ is an integer. Moreover, symmetries of the scatterer may be used to decrease the intervals of Euler angles, over which to average, and hence accelerate the calculation. This feature of ADDA was tested in a recent benchmark study [4].

Other features of ADDA include computation of scattering by a tightly focused Gaussian beams [6], a checkpoint system to allow for long runs on queuing systems that enforce upper limits on wall clock time for execution as is usually the case on massively parallel supercomputers, calculation of radiation forces on each of the dipoles [19], use of rotational symmetry of the scatterer to halve the simulation time, and an extended command line interface. Some other features, such as applicability to anisotropic scatterers and a large set of predefined shapes, are planned to be implemented in the near future.

There are several factors that allow ADDA's performance to compare favorably with other codes, which was shown in a benchmark study by Penttila *et al.* [4]. First of all, the FFTW 3 package that is used automatically adapts itself to optimally perform on any particular hardware. Moreover, ADDA does not perform complete 3D FFT transforms in one run, but decomposes them into a set of 1D transforms with data transposition in between. This allows employing the fact that input data for the forward transform contains many zeros, and only part of the output data of the backward transform is used [8]. Second, we have implemented four different Krylov-space-based iterative solvers, allowing us to choose the most suitable one for a particular application. As is known from the literature [17,20,21] and demonstrated in Section 3, there is not a best iterative solver for DDA. Depending on all details of the scattering problem, any of the methods may outperform the others. Third, dynamic memory allocation and optimized data structures allow all computations, except the FFT, to be performed only for the real (non-void) dipoles and not for the whole computational box. This also decreases ADDA's memory consumption. Moreover, symmetry of the interaction matrix is used to decrease memory required for its Fourier transform. Finally, all float variables in ADDA are represented in double precision. This accelerates convergence in cases when machine precision becomes important. Moreover, basic operations with double-precision numbers can be faster than with single-precision ones on modern processors. This acceleration comes at a cost of increased memory consumption, which is, however, still lower than for other computer codes [4].

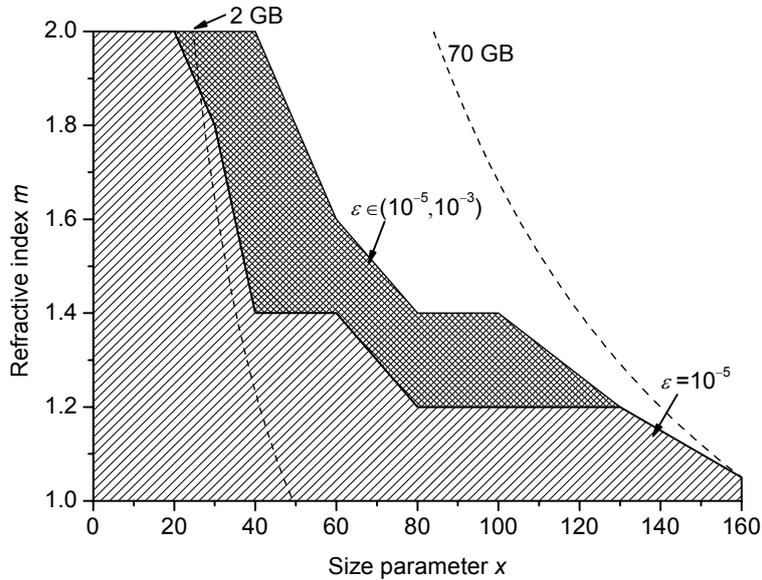

Fig. 1. Current capabilities of the ADDA for spheres with different $x$ and $m$. The striped region corresponds to full convergence and densely hatched region to incomplete convergence. The dashed lines show two levels of memory requirements for the simulation, according to the "rule of thumb" (see main text for explanation).

More information on ADDA can be found in an extensive manual included in the distribution package.

## 3  Numerical simulations

### 3.1  Simulation parameters

In our tests we used ADDA v.0.75, compiled with the Intel C compiler v.9.0 with maximum possible optimizations (default options in ADDA's makefile). All the tests were run on the Dutch compute cluster LISA,[3] using 32 nodes (each dual Intel Xeon 3.4 GHz processor with 4 GB RAM). LDR was used as the most common polarization formulation. We have tried three different iterative solvers: QMR, Bi-CG, and Bi-CGSTAB. For all of them a default stopping criterion $\varepsilon = 10^{-5}$ was used.

---

[3] http://www.sara.nl/userinfo/lisa/description/

Table 1. Parameters of the numerical simulations.

| $m$ | $x$ | $\lambda/md$ | Number of dipoles[a] | Iterative method | Number of iterations |
|---|---|---|---|---|---|
| 1.05 | 20 | 9.6 | $2.6\times10^5$ | Bi-CGSTAB | 6 |
| | 30 | 9.6 | $8.8\times10^5$ | Bi-CGSTAB | 7 |
| | 40 | 9.6 | $2.1\times10^6$ | Bi-CGSTAB | 9 |
| | 60 | 9.6 | $7.1\times10^6$ | Bi-CGSTAB | 14 |
| | 80 | 9.6 | $1.7\times10^7$ | Bi-CGSTAB | 20 |
| | 100 | 9.6 | $3.3\times10^7$ | Bi-CGSTAB | 27 |
| | 130 | 10.3 | $9.0\times10^7$ | Bi-CGSTAB | 40 |
| | 160 | 9.6 | $1.3\times10^8$ | Bi-CGSTAB | 65 |
| 1.2 | 20 | 10.5 | $5.1\times10^5$ | QMR | 86 |
| | 30 | 11.2 | $2.1\times10^6$ | QMR | 223 |
| | 40 | 10.5 | $4.1\times10^6$ | QMR | 598 |
| | 60 | 9.8 | $1.1\times10^7$ | QMR | 2120 |
| | 80 | 10.5 | $3.3\times10^7$ | Bi-CGSTAB | 21748 |
| | 100 | 10.1 | $5.7\times10^7$ | Bi-CGSTAB | 6169 |
| | 130 | 10.3 | $1.3\times10^8$ | Bi-CGSTAB | 29200 |
| 1.4 | 20 | 10.8 | $8.8\times10^5$ | QMR | 1344 |
| | 30 | 10.8 | $3.0\times10^6$ | QMR | 16930 |
| | 40 | 10.8 | $7.1\times10^6$ | QMR | 8164 |
| | 60 | 9.6 | $1.7\times10^7$ | Bi-CG | 127588 |
| 1.6 | 20 | 11.0 | $1.4\times10^6$ | QMR | 8496 |
| | 30 | 10.5 | $4.1\times10^6$ | Bi-CG | 69748 |
| 1.8 | 20 | 11.2 | $2.1\times10^6$ | QMR | 28171 |
| | 30 | 10.2 | $5.5\times10^6$ | Bi-CG | 118383 |
| 2 | 20 | 10.1 | $2.1\times10^6$ | QMR | 58546 |

[a] This is the total number of dipoles in the rectangular computational grid, which is the main factor determining the computation time of one iteration. For spheres the number of dipoles occupied by the scatterer itself is almost two times smaller.

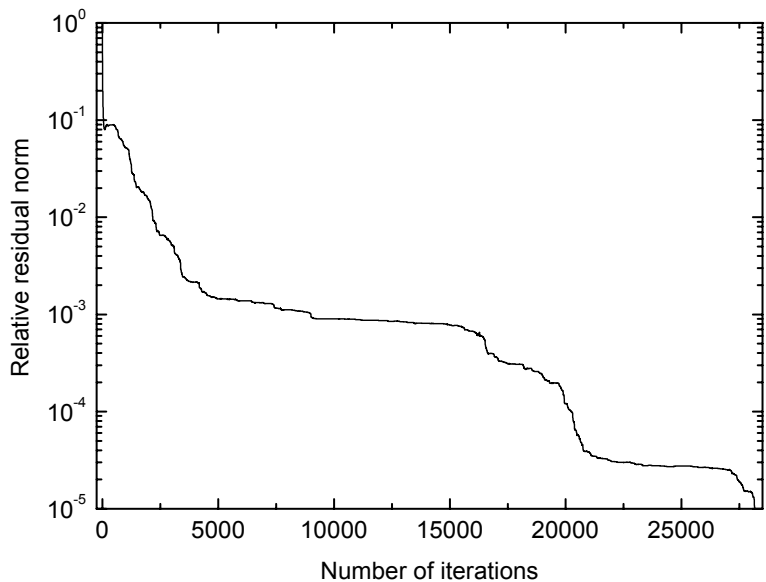

Fig. 2. Convergence of the QMR iterative solver for the sphere with $x = 20$ and $m = 1.8$. The residual as a function of the iteration number is shown. The system of linear equations contains $3\times10^6$ unknowns.

Spheres were used as test objects. Their size parameter $x$ was varied from 20 to 160 and their refractive index $m$ was varied from 1.05 to 2. We limited ourselves to the case of real $m$. The current capabilities of ADDA are shown as a region of the $(x,m)$-plane in Fig. 1. The striped region corresponds to full convergence, the densely hatched region corresponds to those cases where ADDA could not fully converge to the required residual norm, but only to $\varepsilon \in (10^{-5}, 10^{-3})$. Although this incomplete convergence probably affects the final accuracy of the scattering quantities only slightly, we remove such results from further consideration because a separate study is required to quantify this effect (see Section 4). For fully converged results, the errors of scattering quantities due to the numerical convergence are much smaller than the total errors (data not shown).

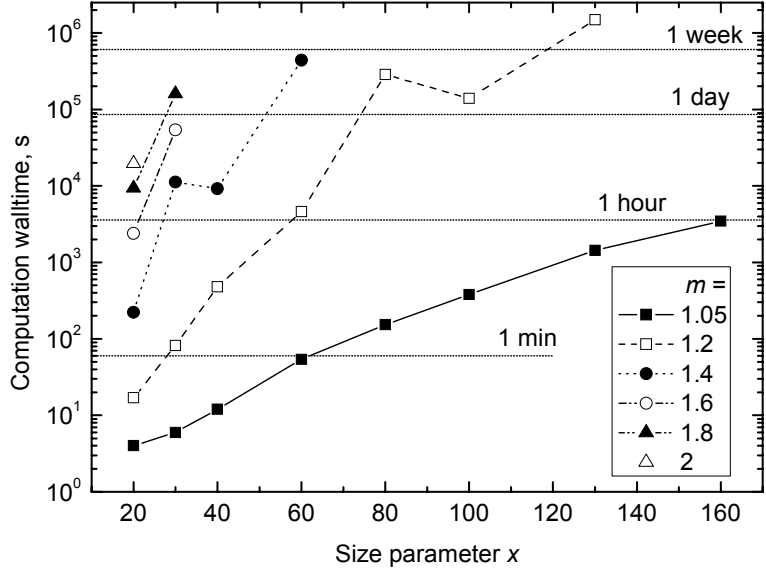

Fig. 3. Total simulation wall clock time (on 64 processors) for spheres with different $x$ and $m$. Time is shown in logarithmic scale. Horizontal dotted lines corresponding to a minute, an hour, a day, and a week are shown for convenience.

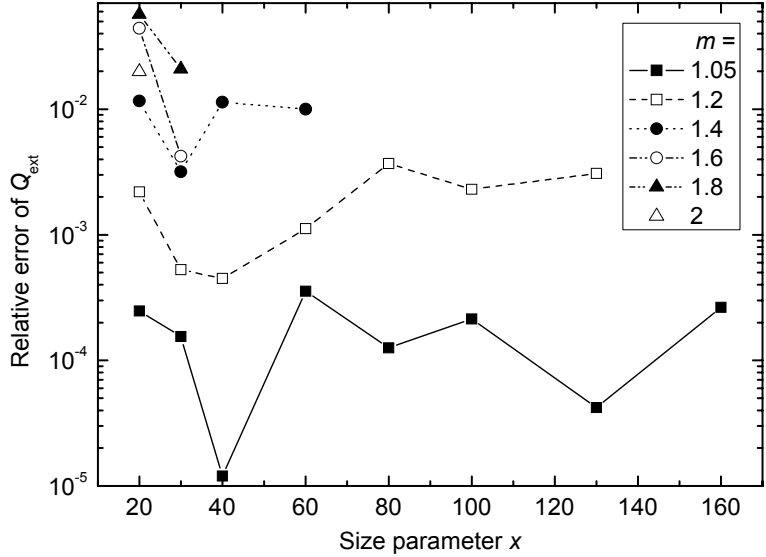

Fig. 4. Relative errors of the extinction efficiency in logarithmic scale for spheres with different $x$ and $m$.

A complete set of ($x$,$m$) pairs, for which ADDA converged, is shown in Table 1. It also shows the number of dipoles per wavelength in the medium ($\lambda/md$ where $d$ is the size of the dipole). We tried to keep it equal to 10 according to the "rule of thumb" as formulated by Draine and Flatau [2]; however, it was slightly different because we varied the size of the dipole grid to optimize the parallel efficiency of ADDA.[4] The total number of dipoles in a rectangular computational grid, shown in Table 1, was varied from $2.6 \times 10^5$ to $1.3 \times 10^8$, it can be approximately determined as $(3.18xm)^3$. Both memory requirements and computation time of one iteration are proportional to this number. Two dashed lines are shown in Fig. 1 to indicate the memory requirements for different $x$ and $m$. They correspond to typical memory of a modern desktop computer (2 GB) and the maximum total memory used in our simulations (70 GB), respectively.

---

[4] The best parallel performance is obtained when grid size divides the number of processors. However, ADDA works with any grid size.

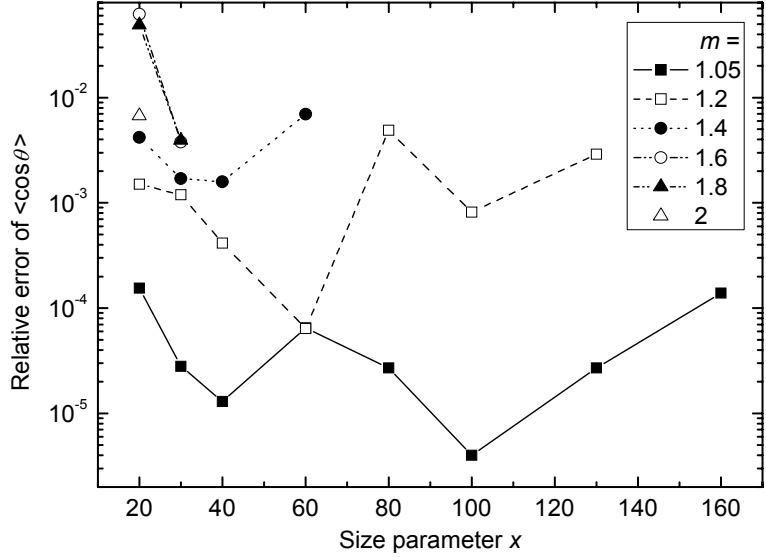

Fig. 5. Same as Fig. 4 but now for the asymmetry parameter.

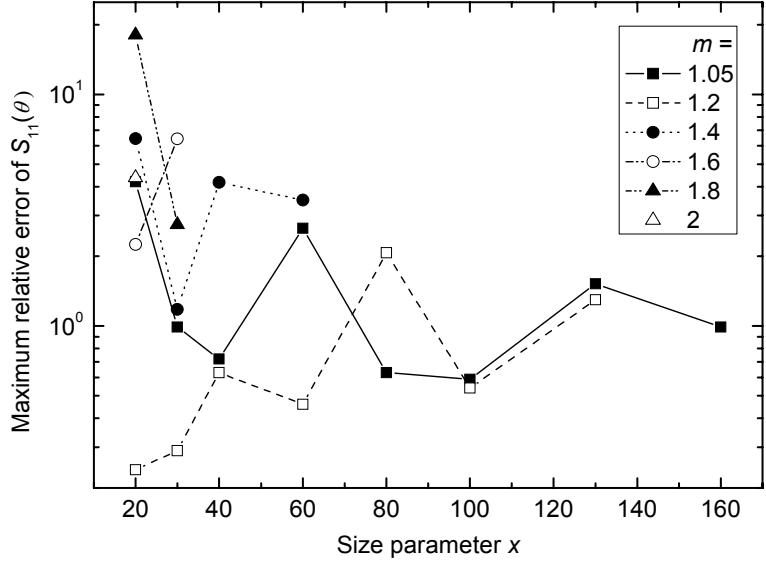

Fig. 6. Maximum relative errors of $S_{11}(\theta)$ in logarithmic scale for spheres with different $x$ and $m$.

For each sphere we computed the extinction efficiency, the asymmetry parameter, and all Mueller matrix elements in one scattering plane, which is a symmetry plane of the cubical discretization of the sphere. Exact results for the same spheres were obtained using the Mie theory [22]. Spherical symmetry was used by ADDA to get all results from calculations for only one polarization state of the incident field. Therefore computation time is a factor of two smaller than for non-symmetric scatterers with the same $x$ and $m$. We employed a volume correction to ensure equal volumes of sphere and its dipole representation [2]. Note, however, that for the very large spheres this correction is extremely small.

## 3.2 Results

Table 1 shows the iterative solver that provided the best performance for each particular case and the number of iterations to achieve convergence. Fig. 2 illustrates one specific example of convergence of the DDA iterative solver. This is QMR applied to the system of $3 \cdot 10^6$ linear equations obtained for the sphere with $x = 20$ and $m = 1.8$. The total simulation wall clock time $t$ for all particles is shown in Fig. 3. Fig. 4 and Fig. 5 show the relative errors of the extinction efficiency $Q_{ext}$ and the asymmetry parameter $<\cos\theta>$ respectively. Maximum - and root-mean-squared (RMS) relative errors of $S_{11}$ over the whole range of scattering angle are shown in Fig. 6 and Fig. 7 respectively. Errors of other non-trivial Mueller matrix elements behave in a similar way (data not shown).

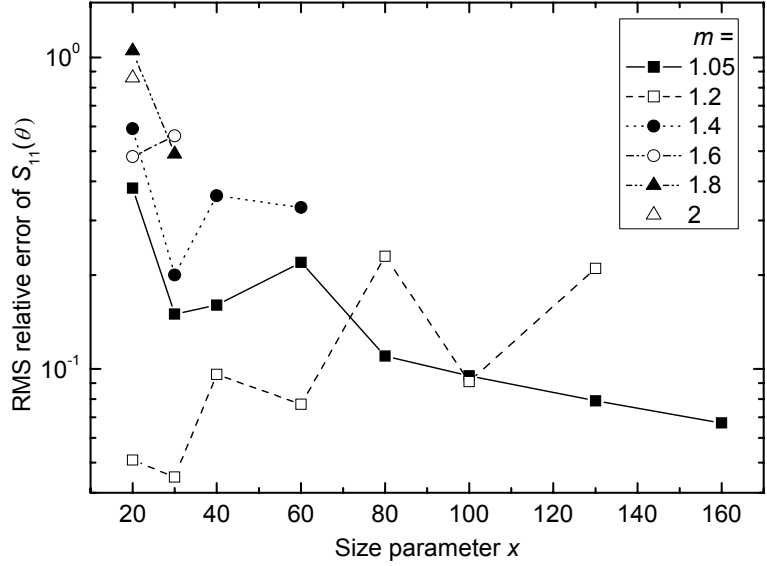

Fig. 7. Same as Fig. 6 but now for RMS relative errors.

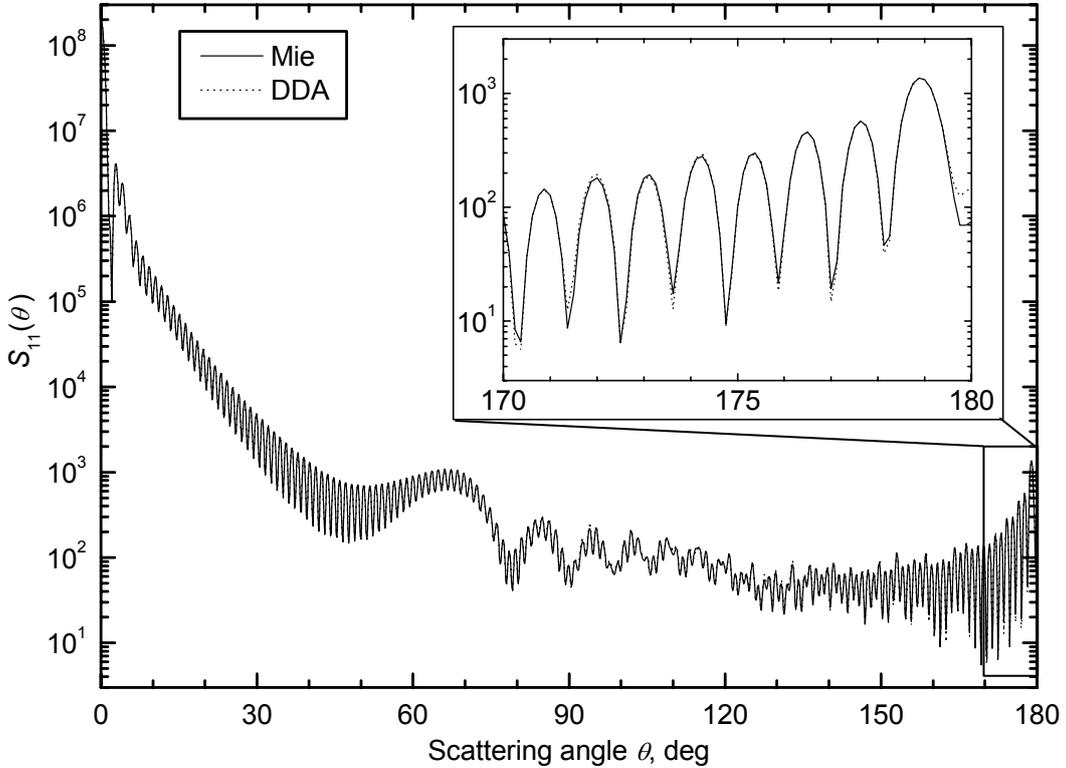

Fig. 8. DDA results (dotted line) of $S_{11}(\theta)$ in logarithmic scale for a sphere with $x = 160$ and $m = 1.05$, compared with the results of the Mie theory (solid line).

DDA results of $S_{11}(\theta)$ for a sphere with $x = 160$ and $m = 1.05$ are compared with the Mie theory in Fig. 8. The inset shows a magnification of the backscattering region. This is, to the best of our knowledge, the largest particle ever simulated with DDA. Fig. 9 and Fig. 10 show the same comparisons but for $x = 60$, $m = 1.4$ and $x = 20$, $m = 2$ respectively.

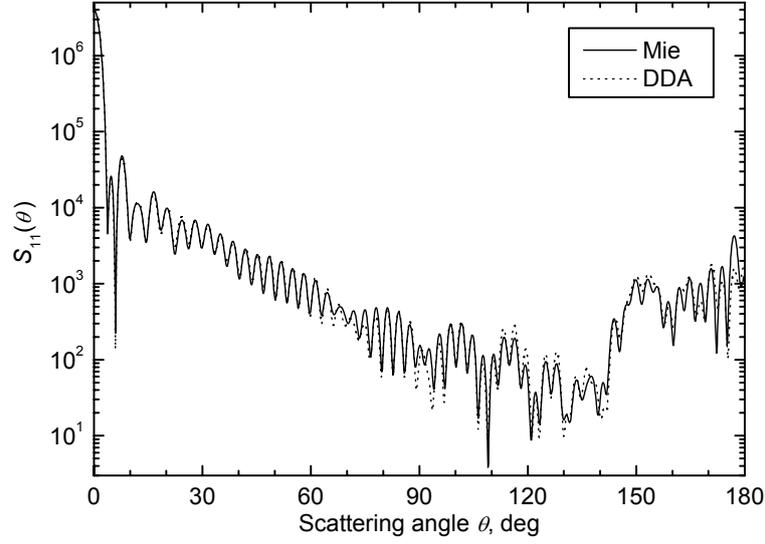

Fig. 9. Same as Fig. 8 but now for $x = 60$ and $m = 1.4$.

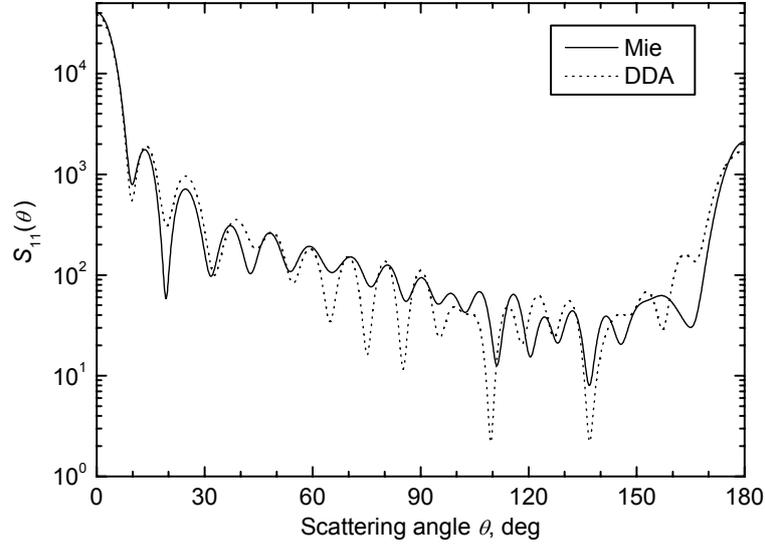

Fig. 10. Same as Fig. 8 but now for $x = 20$ and $m = 2$.

## 4 Discussion

The convergence of the QMR iterative solver shown in Fig. 2, featuring plateaus and steep descents, is in agreement both with its behavior in general [16] and with particular examples of its application to DDA [20,23]. A distinctive feature of this graph compared to the literature data is that the convergence slows down with iteration number, i.e. the logarithm of the residual norm decreases slower than linearly. This is probably due to the large size of the scatterer and loss of numerical precision (see discussion below).

The total computation times $t$ increase steeply both with $x$ and $m$ (Fig. 3). The time is displayed in a logarithmic scale covering a range from 4 seconds to more than 2 weeks. For $m = 1.05$, the increase of $t$ with $x$ is mostly due to the increasing number of dipoles to model the scatterer, since the number of iterations increase at a slower pace (Table 1). For larger $m$ these two effects are comparable, combining into a very unfavorable scaling, which can be approximately described by a power law $t \approx C(m)x^{\alpha(m)}$, where $\alpha > 6$ for $m \geq 1.2$. It should be noted that both the number of iterations and $t$ do not always increase monotonically with $x$. For example for $x = 80$, $m = 1.2$ and $x = 30$, $m = 1.4$ the execution times are unusually high. This may be caused by a large condition number of DDA interaction matrices for these two particular particles. Moreover, when the convergence is slow it may suffer from machine precision, the latter determining the limit of *x* and *m*, for which ADDA will converge at all.

Therefore, current size limitations of the DDA for $m \geq 1.2$ are due to the practically unbearable computation times, and not due to memory requirements.[5] Simulations for larger *m* are far from the memory limit shown in Fig. 1. Moreover, simply using more processors does not solve the problem. Improving numerical performance is required, e.g. dedicated preconditioning of the iterative solver [15]. On the other hand, extension to larger sizes for $m < 1.2$ is feasible if more computer resources are available. This facilitates, for example, simulating scattering of visible light by almost all biological cells in suspension.

The increase of the number of iterations with *m* is a well-known fact [12,17,21,24]; however, there is still no theoretical foundation to describe it in details. Rahola [24] provided theoretical predictions of the dependence of the number of iterations on *m*, valid for scatterers smaller than the wavelength. However, these conclusions are not applicable to the scatterers studied in this manuscript. The general reason for the slowing down of the convergence with increasing *m* is increased interaction between dipoles and, hence, an increased condition number of the interaction matrix. Absorption, if present, should decrease the overall interaction between dipoles in a large scatterer. Therefore, it is expected that convergence for complex refractive indices should be better than for the purely real ones that we consider here. The same was suggested by Budko and Samokhin [25] based on the analysis of the spectrum of the integral scattering operator. However, this proposition is still to be verified by numerical tests.

Another parameter that may greatly affect the computation time is the convergence threshold $\varepsilon$. In this paper it is set to a de-facto default value of $10^{-5}$ [2], which ensures negligibly small numerical errors compared to the model errors. However, in many cases numerical errors are small enough already for $\varepsilon = 10^{-3}$, i.e. the difference of the scattering quantities between simulations with $\varepsilon = 10^{-3}$ and $\varepsilon = 10^{-5}$ is significantly smaller than the difference between the latter and the exact values (data not shown). Fig. 2 shows that QMR for a particular case converges to $\varepsilon = 10^{-3}$ and $\varepsilon = 2 \times 10^{-3}$ three and six times faster respectively than to $\varepsilon = 10^{-5}$. Results for other simulated particles and iterative solvers show similar trends and even larger acceleration with increasing $\varepsilon$ in some cases (data not shown). Therefore, if one can determine an optimum $\varepsilon$ for a particular case, it can decrease the computation time significantly. However, we do not pursue this issue further in this manuscript.

Fig. 4 shows the deterioration of the accuracy of $Q_\text{ext}$ with increasing *m*, while there is no clear dependence on *x* (the only exception is a single result for $m = 2$). Results for $<\cos\theta>$ (Fig. 5) behave in a similar way. These results are in good agreement with results of other researchers for smaller size parameters [2,13,26], both in terms of the errors themselves and their dependence on *m*. To express errors on the angular dependencies of $S_{11}$ we use two integral parameters: the maximum - and RMS relative errors (Fig. 6 and Fig. 7 respectively). Although these parameters are not completely objective, as they are significantly influenced by the values of $S_{11}$ in deep minima, which are completely irrelevant to most real experiments, they do provide a consistent measure of the DDA accuracy. To relate these integral parameters to some other criteria, e.g. visual agreement, three examples are presented in Fig. 8 – Fig. 10. Errors of $S_{11}(\theta)$ show the same tendencies as the integral scattering quantities, except that errors for $m = 1.05$ are relatively large (larger than those for $m = 1.2$ in the range $x \leq 60$) and generally decrease with *x*. This is due to the relative nature of the measured errors and the huge dynamical range of $S_{11}(\theta)$ for small refractive indices (see Fig. 8). Results for smaller size parameters found in the literature [2,26] show a similar increase of errors with *m*: however, the errors themselves are considerably smaller. For instance, maximum relative errors of $S_{11}(\theta)$ for $x<10$ and *m* up to $2.5+1.4i$ are smaller than 0.4. This is due to the general differences between functions $S_{11}(\theta)$ for particles comparable to and much larger than the wavelength. The latter has deeper minima and a larger overall dynamic range. It is important to note that refractive indices as small as 1.05 are rarely used in DDA simulations [26], therefore it is hard to make any definite conclusions concerning the behavior of errors in this case.

---

[5] The boundary value of *m* is not well-defined, as it depends on particular hardware and restrictions on computation time; 1.2 is just a convenient value to guide the reader.

In what follows, the traditional "rule of thumb" [2] is discussed, which states that for $\lambda/md=10$ errors of cross sections and asymmetry parameter are expected to be a few percents, and maximum errors in the angular dependence of $S_{11}$ on the order of 20-30 %. Results for both $Q_{ext}$ and $<\cos\theta>$ do satisfy the "rule of thumb," however this rule does not describe the decrease of errors by two orders of magnitude with decreasing *m*. The latter can be used to cut down the number of dipoles and hence computation time in cases when only integral scattering quantities need to be calculated for small *m*. Relative errors of $S_{11}(\theta)$ are much larger than that predicted by the "rule of thumb," which is due to the fact that the latter was derived based on test simulations for *x* smaller than 10 [2]. See, however, the discussion below on possible changes for complex refractive index and non-spherical shapes. To conclude, the "rule of thumb" has very limited application for the range of *x* and *m* here. More elaborate empirical functions are required to estimate the number of dipoles needed to reach a prescribed accuracy. They will also allow a more realistic estimate of DDA computational complexity, i.e. the computation time needed to reach a certain accuracy of some scattering quantities for particular *x* and *m*. This topic is left for the future study.

The test results shown in this paper are limited to real refractive indices and spherically shaped scatterers. In the following we try to generalize our conclusions to complex refractive index and non-spherical shapes. However, we want to stress that this generalization is speculative, and more numerical tests are clearly needed to verify them. It is expected that accuracies of integral scattering quantities should not change significantly for more general cases. Their accuracy should deteriorate both with increasing real and imaginary parts of the refractive index. The situation for angle-resolved scattering quantities is expected to be different. Large relative errors observed in this paper are due to deep minima that are a consequence of both spherical symmetry and purely real refractive index. It is expected that visual agreement between the DDA results and the exact solution (as shown in Fig. 8 – Fig. 10) should not change significantly for more general cases, however it will result in smaller relative errors, especially for larger *x* and smaller *m*.

## 5  Conclusion

In this paper we present the ADDA, a computer program to simulate light scattering by arbitrarily shaped particles. ADDA can parallelize a single DDA simulation, which allows it not to be limited by the memory of a single computer. Moreover, ADDA is heavily optimized, which allows it to compare favorably with other programs based on DDA when running on a single processor. We showed its capabilities for simulating light scattering by spheres with *x* up to 160 and *m* up to 2. The maximum reachable *x* on a cluster of 64 modern processors decrease rapidly with increasing *m*: it is 160 for $m=1.05$ and only 20-40 (depending on the convergence threshold) for $m=2$. This is mostly due to the slow convergence of the iterative solver leading to practically unbearable computation times. It is expected that larger particle sizes can be reached if *m* has a significant imaginary part.

Errors of both integral and angle-resolved scattering quantities show no systematic dependence on *x*, but generally increase with *m*. Errors of $Q_{ext}$ and $<\cos\theta>$ range from less than 0.01 % to 6 %. Maximum - and RMS relative errors of $S_{11}(\theta)$ are in the ranges 0.2–18 and 0.04–1 respectively. Error predictions of the traditional "rule of thumb" have very limited application in this range of $x$ and $m$: it describes the upper limit of errors of $Q_{\text{ext}}$ and $<\cos\theta>$, however it does not account for the decrease of the errors with $m$.

Currently, the ADDA is capable of simulating light scattering by almost all biological cells in suspension; however, its performance for other cases can be improved. These improvements, left for future work, may include improving the convergence of the iterative solver by preconditioning. It also is desirable to conduct a detailed study of the dependence of the accuracy of the final results on the size of the dipole and convergence thresholds of the iterative solver for different scatterers. Such a study should result in a reduction of the computation time and provide a realistic estimate of DDA complexity over a wide range of $x$ and $m$.

## Acknowledgements

We thank Gorden Videen for critically reading the manuscript and anonymous reviewer for valuable comments. Our research is supported by Siberian Branch of the Russian Academy of Sciences through the grant 2006-03.